{
\documentclass[onecolumn,showpacs,amsmath,amssymb,aps]{revtex4}

\usepackage{graphicx}
\usepackage{dcolumn}
\usepackage{bm}

\usepackage{amsmath}
\usepackage{amssymb}
\usepackage{url}
\usepackage{subfigure}
\usepackage{algorithm}
\usepackage{algorithmic}
\usepackage{enumerate}
\usepackage{slashbox}
\usepackage{setspace}
\usepackage{ulem}

\begin{document}

\preprint{APS/123-QED}

\title{On approximate equivalence of modularity, D and non-negative matrix factorization}

\author{Zhenhai Chang}
\affiliation{School of Statistics and Mathematics, Central University of Finance and Economics}

\author{Hui-Min Cheng}%
\affiliation{School of Statistics and Mathematics, Central University of Finance and Economics}

\author{Chao Yan}%
\affiliation{School of Statistics and Mathematics, Central University of Finance and Economics}

\author{Xianjun Yin}%
\affiliation{School of Statistics and Mathematics, Central University of Finance and Economics}

\author{Zhong-Yuan Zhang}%
\email{zhyuanzh@gmail.com}
\affiliation{School of Statistics and Mathematics, Central University of Finance and Economics}

\date{\today}

\begin{abstract}
Community structures detection is one of the fundamental problems in complex network analysis
towards understanding the topology structures of the network and the functions of it. Nonnegative
matrix factorization (NMF) is a widely used method for community detection, and modularity Q and
modularity density D are criteria to evaluate the quality of community structures. In this paper, we
establish the connections between Q, D and NMF for the first time. Q maximization can be
approximately reformulated under the framework of NMF with Frobenius norm, especially when $n$ is
large, and D maximization can also be reformulated under the framework of NMF. Q minimization can
be reformulated under the framework of NMF with Kullback-Leibler divergence. We propose new
methods for community structures detection based on the above findings, and the experimental results
on synthetic networks demonstrate their effectiveness.
\end{abstract}

\pacs{Valid PACS appear here}
\maketitle

\section{\label{Introduction}Introduction}
Many real-world systems can be expressed as complex networks, where nodes and edges represent elements in the system and the relations among them, respectively, such as social networks \cite{1Snijders2011Statistical}, biological networks \cite{2zhong2016an}, information networks \cite{3albert1999internet}, etc. One of the important statistical properties of complex networks is community structure \cite{girvan2002community}. Although there is not an accurate definition, it is widely accepted that community in the network is set of nodes, which are densely interconnected but loosely connected with the rest \cite{girvan2002community, 5fortunato2009community, newman2003the}. Community detection can help us better understand the topological structures of the network from the aspect of meso-level and network's functions, such as epidemic spreading and information diffusion \cite{guimera2005functional,5fortunato2009community}.

To quantitatively describe community structures, modularity $Q$ \cite{31Newman2004Finding} is proposed and has been extensively studied. The basic idea of modularity is very similar with hypothesis testing: given a network partition, it compares the fraction of links inside each module of the partition with the so-called null model, i.e., the expected fraction of links inside the corresponding module in degree preserving counterpart, and sums the differences over all of the modules of the partition \cite{Newman2006modularity,Lancichinetti2011Limits,Hofstad2009Random}. A higher modularity value indicates that the partition is reasonable and meaningful, or one can say it has statistical significance. Modularity can be not only optimized directly for community detection, but also used for evaluation of community structures detected by other methods. However, modularity-optimization strategy has a resolution limit \cite{32Fortunato2006Resolution, 33Good2010Performance}. Modularity maximization usually tends to merge small communities into large ones in network. To overcome this problem, \cite{36Li2008Quantitative} proposed a function called modularity density (also called $D$) to evaluate the partition of a network into communities.

Besides modularity, there are also other cost functions that can be optimized over all possible network partitions for community detection, such as nonnegative matrix factorization (NMF) \cite{30Lee1999Learning,Lee2001Algorithms}.  In recent years, NMF has been successfully applied to a range of community structures detection including nonoverlapping communities, overlapping ones and bipartite ones \cite{cao2013identifying, Zhang2013Enhanced,zhang2015community}. Furthermore, by adjusting objective functions, many community detection methods can be reformulated under the framework of NMF. For example,  maximum likelihood of stochastic block models (SBM), spectral clustering and probabilistic latent semantic indexing can be reformulated by using the objective functions of NMF, respectively \cite{18Zhang2016On, 19Ding2005On, 20Ding2008On, 21Devarajan2015A}.

It has been proven that modularity maximization and likelihood maximization of stochastic block models are equivalent under appropriate conditions\cite{22Newman2016Community}, and the likelihood maximization of stochastic block model can be reformulated under the framework of nonnegative matrix factorization \cite{18Zhang2016On}. These results naturally lead to the next interesting question: \textit{Is modularity maximization equivalent with NMF?}

Indeed, although both modularity optimization and NMF are widely used community detection methods, the relations of them have not been established. In this paper, the relations of modularity, $D$ and NMF are discussed. Despite the model relations, the algorithms employed by $Q$, $D$ and NMF are different. We propose new multiplicative update rules for NMF, and demonstrate its effectiveness on synthetic networks.

The implications of the work are three folds: 1. There is a general framework for $Q$ and $D$ optimization. 2. It should be cautious when using $Q$ and $D$ for evaluation of detected communities by different methods, especially by NMF, since they are (approximately) equivalent. 3. The relations between $Q$, $D$, NMF and stochastic block model shed light on designing more effective algorithms for community detection.

The rest of the paper is organized as follows. In Sect.II, the equivalence of modularity maximization and minimizing the object function of NMF with Frobenius norm is proven. In Sect.III, the equivalence of $D$ maximization and minimizing the object function of NMF with Frobenius norm is proven. In Sect.IV, modularity minimization and minimizing the object function of NMF with Kullback-Leibler divergence is proven. In Sect.V, we empirically demonstrate the relations between $Q$ and NMF, and $D$ and NMF. Furthermore, the algorithms employed by $Q$, $D$ and NMF are compared. Finally, Sect.VI concludes.

\section{\label{QNMFF}Approximate Equivalence of modularity maximization and NMF with Frobenius norm}
Modularity optimization is a widely used method where the benefit function $Q$ is defined to measure the quality of divisions of a network into communities. Many modularity optimization schemes have been proposed \cite{Newman2004Fast, Duch2005community, Guimera2004modularity}. In this section, we prove that modularity maximization is approximately equivalent to minimizing the object function of NMF with Frobenius norm.

Suppose that network $G=(V, E)$ can be divided into $c$ communities $\{V_1, V_2, \cdots, V_c\}$ satisfying the following conditions:
\[
V = \bigcup\limits_{r = 1}^c{V_r } ,\;V_r  \cap V_q  = \varnothing,\;\forall r \ne q,
\]
where $V=\lbrace 1,2,\cdots,n \rbrace $ is the set of nodes and $E$ is the set of edges.

Modularity is defined as \cite{31Newman2004Finding}
\[
Q = \frac{1}{{2m}}\sum\limits_{i,j = 1}^n {(A_{ij}  - \frac{{k_i k_j }}{{2m}})\delta (g_i ,g_j )} ,
\]
where $m$ is the total number of edges in the network, $\delta ( \cdot , \cdot )$ is the Kronecker delta, $ g_{i} $ is the community to which node $i$ belongs, $ g_{i} \in \{1, 2, \cdots, c \} $, $k_i$ is the degree of node $i$, $A=(A_{ij})_{n \times n}$ is the adjacency matrix of $G$ with entries $A_{ij}  =1$ if there is an edge between nodes $i$ and $j$ and $0$ otherwise.

Let
\[
s_{ir}  = \left\{ \begin{array}{l}
1, \mbox{if} \;\; g_i = r, \\
0, \mbox{otherwise.} \\
\end{array} \right.
\]

Then, one has
\[
 \delta (g_i ,g_j ){\rm{ = }}\sum\limits_{r = 1}^c {s_{ir} s_{jr} }, \sum\limits_{r = 1}^c {s_{ir} }  = 1, Tr(S^T S) = n.
\]
where $ S = (s_{ir} )_{n \times c} $.  For a given network, let $Q^*  =  (2m)^2 Q $, and one has \cite{28White2005A}
\[
\begin{array}{l}
Q^* = \;\sum\limits_{i,j = 1}^n {(2mA_{ij}  - k_i k_j )\delta (g_i ,g_j )}  \\
\;\;\;\;\; = \sum\limits_{i,j = 1}^n {(2mA_{ij}  - k_i k_j )\sum\limits_{r = 1}^c {s_{ir} s_{jr} } }  \\  \vspace{1ex}
\;\;\;\;\; = \sum\limits_{r = 1}^c {\left[ {\sum\limits_{i,j = 1}^n {\left( {2mA_{ij} s_{ir} s_{jr}  - k_i k_j s_{ir} s_{jr} } \right)} } \right]}  \\  \vspace{1ex}
\;\;\;\;\; = \sum\limits_{r = 1}^c {\left[ {\sum\limits_{i,j = 1}^n {2mA_{ij} s_{ir} s_{jr}  - \sum\limits_{i = 1}^n {k_i s_{ir} } \sum\limits_{j = 1}^n {k_j s_{jr} } } } \right]}  \\  \vspace{1ex}
\;\;\;\;\; = \sum\limits_{r = 1}^c {\left[ {\sum\limits_{i,j = 1}^n {2mA_{ij} s_{ir} s_{jr}  - \left( {\sum\limits_{i = 1}^n {k_i s_{ir} } } \right)^2 } } \right]}  \\ \vspace{1ex}
\;\;\;\;\;  = Tr\left( {S^T (W - Z)S} \right), \\
\end{array} \eqno(1)
\]
where $W=2mA$, $ b=(k_1, k_2, \cdots, k_n)^T, Z=bb^T$. Actually, $\arg \max Q$ and $\arg \max Q^*$ with respect to $S$ are identical, since the number of edges $m$ is constant. Hence,
\[
\mathop {\arg \max }\limits_{S^T S = \Delta ,\;S \ge 0} Q =  \mathop {\arg \max }\limits_{S^T S = \Delta ,\;S \ge 0} \left\{ {Tr\left( {S^T (W - Z)S} \right)} \right\}, \eqno(2)
\]
where $\Delta=diag(n_1, n_2, \cdots, n_c)$, and $n_r$ is the number of nodes in community $V_r$.

Since the values of $S$ are discrete, and the number of possible divisions of a network is exponentially large, one normally turns to approximate optimization methods \cite{22Newman2016Community}, and $S$ can be obtained by the following two steps.

Firstly, by relaxing the constraints on $S$ from binary to non-negative, we can derive the solution $S$ for modularity optimization \cite{28White2005A}. Consider Lagrange multiplicator method and write
\[
J(S) = Tr\left( {S^T (W - Z)S} \right) + (S^T S - \Delta )\Lambda, \eqno(3)
\]
where $\Lambda$ is the diagonal matrix of Lagrangian multipliers. Let $ \displaystyle \frac{{\partial J}}{{\partial s_{ir} }} = 0$, one has
\[
L_Q S = S\Lambda, \eqno(4)
\]
where $L_Q=Z - W$.

Furthermore, when $n$ is sufficiently large, Eq.(4) can be approximated as \cite{28White2005A}
\[
L_Q^* S = S\Lambda, \eqno(5)
\]
where $ \displaystyle L_Q^*= \frac{1}{{n^2 }}E - \frac{1}{n}W^*$, and $E$ is a matrix whose entries are one, $W^* = B^{ - 1} A, B=diag(k_1, k_2, \cdots, k_n)$. Obviously, as $ n \rightarrow \infty $, $ \displaystyle \frac{1}{{n^2 }}E $ approach $0$ much faster than $ \displaystyle \frac{1}{n}W^* $ \cite{28White2005A}. As a result, the first term $ \displaystyle \frac{1}{{n^2 }}E $ can be neglected in determining the eigenspace of the matrix when $n$ is sufficiently large \cite{28White2005A}. Note that the constant $ \displaystyle -\frac{1}{n}$ do not affect the resulting eigenspace, hence Eq.(5) can be represented as follows.
\[
W^*S = S\Lambda. \eqno(6)
\]

From Eqs.(4-6), for sufficiently large values of $n$,  $ \mathop {\arg \max }\limits_{S^T S = \Delta ,\;S \ge 0} Q$ is approximately equivalent to the solution of equation $W^*S = S\Lambda$ (i.e. Eq.(6)).

Write $ J_1 (S) = Tr\left( {S^T W^* S} \right) + (S^T S - \Delta )\Lambda $, let $ \displaystyle \frac{{\partial J_1}}{{\partial s_{ir} }} = 0$, we can derive $W^*S = S\Lambda$ (i.e. Eq.(6)). Thus, the solution $S$ of Eq.(6) is the solution of optimizing $Tr\left( {S^T W^* S} \right)$. So, the solution $S$ of  optimizing $Q$ can be approximated by the solution $S$ of  optimizing $Tr\left( {S^T W^* S} \right)$, that is
\[
\mathop {\arg \max }\limits_{S^T S = \Delta ,\;S \ge 0} Q \approx \mathop {\arg \max }\limits_{S^T S = \Delta ,\;S \ge 0} \left\{ {Tr\left( {S^T W^* S} \right)} \right\}{\rm{ = }}\mathop {\arg \min }\limits_{S^T S = \Delta ,\;S \ge 0} \left\{ {\left\| {W^*  - SS^T } \right\|_F^2 } \right\}. \eqno(7)
\]

Eq.(7) means that maximizing Q and minimizing the object function of NMF with Frobenius norm are approximately equivalent especially when $n$ is large.

\section{\label{DNMFF}Equivalence of D maximization and NMF with Frobenius norm}
 Although $Q$ is widely used as a measure of the quality of the detected community structures, it has several drawbacks \cite{Bagrow2012Communities, Zhang2009Modularity, 33Good2010Performance}. Specifically, modularity $Q$ has the problem of resolution limit  \cite{32Fortunato2006Resolution, 33Good2010Performance}. One reason for this drawback is that modularity function $Q$ depends on the total number of edges so that modularity optimization method is difficult to find small communities in large networks \cite{32Fortunato2006Resolution}. To overcome this issue, a measure called modularity density ($D$) has been proposed \cite{36Li2008Quantitative}. $D$ depends on the size of communities instead of the total number of edges. In practice, $D$ maximization can find small communities that modularity optimization can not find \cite{36Li2008Quantitative}.

In this section, we prove that $D$ maximization is equivalent to minimizing the object function of NMF with Frobenius norm.

To simplify notations, we define $L(V_r ,V_q )$ to be the number of links between communities $V_r$ and $V_q$:
\[
L(V_r ,V_q ) = \sum\limits_{i \in V_r ,j \in V_q } {A_{ij} }.
\]

The degree of a set $V_r$ is the sum of degrees of  nodes in community $V_r$:
\[
d_r \buildrel \Delta \over =  \sum\limits_{i \in V_r} {k_{i}}  = L(V_r ,V).
\]
where $k_i  = \sum\limits_{j = 1}^n {A_{ij} } $ is the degree of node $i$.

Then, we denote $L(V_r ,V_r )$ by $d_r^{stay}$, which measures how many links stay within $V_r$ itself, and denote $L(V_r ,V \setminus V_r )$ by $d_r^{escape}$, which measures how many links escape from $V_r$. One has
\[
d_r^{stay} {\rm{ + }}d_r^{escape}  = d_r. \eqno(8)
\]

$D$ is defined as  \cite{36Li2008Quantitative}:
\[
\begin{array}{l}
\displaystyle D = \sum\limits_{r = 1}^c {\left( {\frac{{d_r^{stay}  - d_r^{escape} }}{{n_r }}} \right)},  \\
\end{array} \eqno(9)
\]
where $n_r$ is the number of nodes in community $V_r$.

 Note that
\[
s_{ir}  = \left\{ \begin{array}{l}
1, \mbox{if} \;\; g_i = r, \\
0, \mbox{otherwise.} \\
\end{array} \right.
\]
and
\[
S = \left( {s_{ir} } \right)_{n \times c}  = \left( {s_1 ,s_2 , \cdots ,s_c } \right),
\]
where $ s_r  = \left( {s_{1r} ,s_{2r} , \cdots ,s_{nr} } \right)^T. $
Let $ B = diag(k_1 ,k_2 , \cdots ,k_n ), $  where $k_i  = \sum\limits_{j = 1}^n {A_{ij} } $ is the degree of node $i$, one has
\[
d_r^{stay}  = s_r^T As_r ,\;d_r  = s_r^T Bs_r. \eqno(10)
\]

Since $d_r^{escape}  = d_r  - d_r^{stay}  = s_r^T Bs_r  - s_r^T As_r$,  one has,
\[
\begin{array}{l}
\displaystyle D  = \sum\limits_{r = 1}^c {\left( {\frac{{d_r^{stay}  - d_r^{escape} }}{{n_r }}} \right)}  \\ \vspace{1ex}
\displaystyle \;\;\;\;= \sum\limits_{r = 1}^c {\left( {\frac{{s_r^T As_r  - s_r^T Bs_r  + s_r^T As_r }}{{n_r }}} \right)}  \\  \vspace{1ex}
\displaystyle \;\;\;\; = \sum\limits_{r = 1}^c {\left( {\frac{{s_r^T (2A - B)s_r }}{{n_r }}} \right)} . \\
\end{array} \eqno(11)
\]

Let
\[
H = \left( {H_{ir} } \right)_{n \times c}  \buildrel \Delta \over = \left( {h_1 ,h_2 , \cdots ,h_c } \right), \eqno(12)
\]

where $ \displaystyle h_r  = \frac{{s_r }}{{\sqrt {n_r } }} = \left( {s_{1r} ,s_{2r} , \cdots ,s_{nr} } \right)^T /\sqrt {n_r } .$ One has
\[
D  = \sum\limits_{r = 1}^c {\left( {h_r^T (2A - B)h_r } \right) = Tr\left( {H^T W_1H} \right)} , \eqno(13)
\]
where $W_1=2A-B.$ Let
\[
W_1=\sigma I+2A-B, \eqno(14)
\]
where $I$ is the $n \times n$ identity matrix, $ \sigma $ is a sufficiently large number and independent of $A$ and $B$ so that $W_1$ is non-negative. Then, one has
\[
\begin{array}{l}
\mathop {\arg \max }\limits_{H^T H = I,H \ge 0} \left\{ {D} \right\} = \mathop {\arg \max }\limits_{H^T H = I,H \ge 0} \left\{ {Tr\left( {H^T W_1H} \right)} \right\} \\  \vspace{1ex}
\;\;\;\;\;\;\;\;\;\;\;\;\;\;\;\;\;\;\;\;\;\;\;\;\;\; = \mathop {\arg \min }\limits_{H^T H = I,H \ge 0} \left\{ { - 2Tr\left( {H^T W_1H} \right)} \right\} \\  \vspace{1ex}
\;\;\;\;\;\;\;\;\;\;\;\;\;\;\;\;\;\;\;\;\;\;\;\;\;\; = \mathop {\arg \min }\limits_{H^T H = I,H \ge 0} \left\{ {\left\| W_1 \right\|^2_F  - 2Tr\left( {H^T W_1H} \right) + \left\| {H^T H} \right\|^2_F } \right\} \\ \vspace{1ex}
\;\;\;\;\;\;\;\;\;\;\;\;\;\;\;\;\;\;\;\;\;\;\;\;\;\;= \mathop {\arg \min }\limits_{H^T H = I,H \ge 0} \left\{ {\left\| {W_1 - HH^T} \right\|^2_F } \right\}, \\
\end{array} \eqno(15)
\]

Eq.(15) means that $D$ maximization and minimization of the object function of NMF with Frobenius norm are equivalent.

\section{Equivalence of modularity minimization and NMF with KL divergence}
 Modularity maximization can be used for detecting assortative network structures, and on the other hand, modularity minimization usually reveals disassortative structures \cite{22Newman2016Community}. In Sect.II, we have discussed the relationship between modularity maximization and NMF with Frobenius norm. In this section, we will show that, under certain conditions, modularity minimization is equivalent to minimizing the object function of NMF with KL divergence.

For modularity function $ \displaystyle Q = \frac{1}{{2m}}\sum\limits_{i,j = 1}^n {(A_{ij}  - P_{ij})\delta (g_i ,g_j )} $,
where $ P_{ij} $ is the expected number of edges between nodes $i$ and $j$, if we suppose $P_{ij}$ is a constant (e.g., $ \displaystyle P_{ij}=\frac{16\times16}{2m}$ on famous GN networks), then $\log(P_{ij})$ is also a constant, one has
\[
\begin{array}{l}
\displaystyle Q^{\#} = \left[ {\log (P_{ij} )} \right] \cdot Q \\
\displaystyle \;\;\;\;\;\; = \frac{1}{{2m}}\sum\limits_{i,j = 1}^n {(A_{ij} \log (P_{ij} ) - P_{ij} \log (P_{ij} ))\delta (g_i ,g_j )}  \\  \vspace{1ex}
\displaystyle \;\;\;\;\;\; = \frac{1}{{2m}}\sum\limits_{i,j = 1}^n {\left[ {A_{ij} \log (P_{ij} ) - P_{ij} } \right]\delta (g_i ,g_j )}  + \frac{1}{{2m}}\sum\limits_{i,j = 1}^n {\left[ {P_{ij}  - P_{ij} \log (P_{ij} )} \right]\delta (g_i ,g_j )}.  \\
\end{array} \eqno(16)
\]

Note that $\delta (g_i ,g_j ){\rm{ = }}\sum\limits_{r = 1}^c {s_{ir} s_{jr} }$, one has
\[
\begin{array}{l}
\displaystyle Q^{\#}  = \frac{1}{{2m}}\sum\limits_{i,j = 1}^n {\left[ {A_{ij} \sum\limits_{r = 1}^c {\log (S_{ir} S_{jr} P_{ij} )}  - \sum\limits_{r = 1}^c {S_{ir} S_{jr} P_{ij} } } \right]}  + \frac{1}{{2m}}\sum\limits_{i,j = 1}^n {\sum\limits_{r = 1}^c {S_{ir} S_{jr} } \sigma_1 I}  \\  \vspace{1ex}
\displaystyle \;\;\;\;\;\; = \frac{1}{{2m}}\sum\limits_{i,j = 1}^n {\left[ {A_{ij} \log P_{ij}  \odot (SS^T )_{ij}  - P_{ij}  \odot (SS^T )_{ij} } \right]}  + \frac{1}{{2m}}Tr(S^T \sigma_1 IS) \\  \vspace{1ex}
\displaystyle \;\;\;\;\;\; =  - \frac{1}{{2m}}\sum\limits_{i,j = 1}^n {\left[ {A_{ij} \log \frac{1}{{P_{ij}  \odot (SS^T )_{ij} }} + P_{ij}  \odot (SS^T )_{ij} } \right]}  + \frac{{n\sigma_1 }}{{2m}}, \\
\end{array} \eqno(17)
\]
where $ \sigma_1 = {P_{ij}  - P_{ij} \log (P_{ij} )} $ is a positive real number, and $ \odot $ represents the dot product of two matrices with the same dimensions. Note that $ \log (P_{ij} ) < 0 $, one has
\[
\begin{array}{l}
\mathop {\arg \min }\limits_{S^T S  = \Delta ,\;S \ge 0} Q{\rm{ = }}\mathop {\arg \max }\limits_{S^T S = \Delta ,\;S \ge 0} Q^\#   \\  \vspace{1.5ex}
\displaystyle \;\;\;\;\;\;\;\;\;\;\;\;\;\;\;\;\;\;\;\;\;\;= \mathop {\arg \max }\limits_{S^T S = \Delta ,S \ge 0} \left\{ { - \frac{1}{{2m}}\sum\limits_{i,j = 1}^n {\left[ {A_{ij} \log \frac{{A_{ij} }}{{P_{ij}  \odot (SS^T )_{ij} }} - A_{ij}  + P_{ij}  \odot (SS^T )_{ij} } \right]} } \right\} \\  \vspace{1.5ex}
\displaystyle \;\;\;\;\;\;\;\;\;\;\;\;\;\;\;\;\;\;\;\;\;\;= \mathop {\arg \min }\limits_{S^T S = \Delta ,S \ge 0} \left\{ {\frac{1}{{2m}}\sum\limits_{i,j = 1}^n {\left[ {A_{ij} \log \frac{{A_{ij} }}{{P_{ij}  \odot (SS^T )_{ij} }} - A_{ij}  + P_{ij}  \odot (SS^T )_{ij} } \right]} } \right\}. \\
\end{array} \eqno(18)
\]

Eq.(18) means that minimization Q is equivalent to minimizing the object function of NMF with KL divergence.

Similarly, for $Q^{RB} (\gamma )$ \cite{34Reichardt2006Statistical}
\[
Q^{RB} (\gamma ) = \frac{1}{{2m}}\sum\limits_{i,j = 1}^n {(A_{ij}  - \gamma \frac{{k_i k_j }}{{2m}})\delta (g_i ,g_j )}, \eqno(19)
\]
and $Q^{AFG} (r)$ \cite{35Arenas2008Analysis}
\[
Q^{AFG} (r) = \frac{1}{{2\tilde m}}\sum\limits_{i,j = 1}^n {(\tilde A_{ij}  - \frac{{\tilde k_i \tilde k_j }}{{2\tilde m}})\delta (g_i ,g_j )}, \eqno(20)
\]
where
\[
\begin{array}{l}
2\tilde m = 2m + nr, \\
\tilde k_i  = k_i  + r, \\
\tilde A_{ij}  = \left\{ \begin{array}{l}
A_{ij} ,\; \mbox{if}\;i \ne j, \\
r,\;\;\; \mbox{if}\;i = j. \\
\end{array} \right. \\
\end{array}
\]
one has
\[
\begin{array}{l}
\displaystyle \mathop {\arg \min }\limits_{S^T S = \Delta ,S \ge 0} Q^{RB}  = \mathop {\arg \min }\limits_{S^T S = \Delta ,S \ge 0} \left\{ {\frac{1}{{2m}}\sum\limits_{i,j = 1}^n {\left[ {A_{ij} \log \frac{{A_{ij} }}{{P_{ij}^*  \odot (SS^T )_{ij} }} - A_{ij}  + P_{ij}^*  \odot (SS^T )_{ij} } \right]} } \right\}, \\
\end{array} \eqno(21)
\]
and
\[
\begin{array}{l}
\displaystyle \mathop {\arg \min }\limits_{S^T S = \Delta ,S \ge 0} Q^{AFG}  = \mathop {\arg \min }\limits_{S^T S = \Delta ,S \ge 0} \left\{ {\frac{1}{{2\tilde m}}\sum\limits_{i,j = 1}^n {\left[ {\tilde A_{ij} \log \frac{{\tilde A_{ij} }}{{\tilde P_{ij}  \odot (SS^T )_{ij} }} - \tilde A_{ij}  + \tilde P_{ij}  \odot (SS^T )_{ij} } \right]} } \right\}, \\
\end{array} \eqno(22)
\]
where
\begin{spacing}{1.5}
\[
\begin{array}{l}
P_{ij}^*  = \gamma P_{ij} = {\rm{constant}}, \log (P_{ij}^* ) = {\rm{constant}}, r > 0, \\
\tilde A_{ij}=A_{ij} (i \neq j),  \tilde A_{ii}=r, \\ 
\tilde k_i  = k_i  + r = {\rm{constant}}, \\  
\displaystyle \tilde P_{ij}  = \frac{{\tilde k_i \tilde k_j }}{{2 \tilde m}} = {\rm{constant}},\; \\  
\log (\tilde P_{ij} ) = {\rm{constant}}.
\end{array} \eqno(23)
\]
\end{spacing}
So, Eq.(21) means minimizing $ Q^{RB} (\gamma ) $ with respect to $S$ is equivalent to
\[
\left\{ \begin{array}{l}
\displaystyle \mathop {\min }\limits_S \left\{ {\frac{1}{{2m}}\sum\limits_{i,j = 1}^n {\left[ {A_{ij} \log \frac{{A_{ij} }}{{P_{ij}  \odot (SS^T )_{ij} }} - A_{ij}  + P_{ij}  \odot (SS^T )_{ij} } \right]} } \right\}, \\
\;\;\;s.t.\;\;\sum\limits_{r = 1}^c {s_{ir} }  = 1,\;\;\;i = 1,2, \cdots ,n. \\
\;\;\;\;\;\;\;\;\;\;s_{ir}  \ge 0,\;\;\;i = 1,2, \cdots ,n;\;\;r = 1,2, \cdots ,c. \\
\end{array} \right. \eqno(24)
\]
and  Eq.(22) means minimizing  $ Q^{AFG} (r) $ with respect to $S$ is equivalent to
\[
\left\{ \begin{array}{l}
\displaystyle \mathop {\min }\limits_S \left\{ {\frac{1}{{2\tilde m}}\sum\limits_{i,j = 1}^n {\left[ {\tilde A_{ij} \log \frac{{\tilde A_{ij} }}{{\tilde P_{ij}  \odot (SS^T )_{ij} }} - \tilde A_{ij}  + \tilde P_{ij}  \odot (SS^T )_{ij} } \right]} } \right\}, \\
\;\;\;s.t.\;\;\sum\limits_{r = 1}^c {s_{ir} }  = 1,\;\;\;i = 1,2, \cdots ,n. \\
\;\;\;\;\;\;\;\;\;\;s_{ir}  \ge 0,\;\;\;i = 1,2, \cdots ,n;\;\;r = 1,2, \cdots ,c. \\
\end{array} \right. \eqno(25)
\]

\section{Experimental results}
 In this section, we evaluate the theoretical equivalence proved in Sect.II, Sect.III and Sect. IV using SBM networks (i.e., networks generated using stochastic block model \cite{Holland1983stochastic}) and LFR networks \cite{Lancichinetti2009Benchmarks}. Furthermore, although both $Q$ and $D$ are reformulated under the framework of NMF, their algorithms are different. To demonstrate the effectiveness of the algorithms, we compare them on GN and LFR networks.

 \subsection{Description of data sets}
1. SBM network and its special case: GN network \cite{girvan2002community}

In the stochastic block model (SBM) network, nodes are assigned to $c$ communities with probabilities $\pi= \{ \pi_1, \pi_2, \cdots, \pi_c \}$, and edges are placed randomly and independently between node pairs with probabilities $ \theta_{rs} (r, s=1,2, \cdots, c)$ that depend only on the group memberships of the nodes. The framework is flexible such that many different kinds of networks can be produced.

In this paper, the between-community edge probability $ \theta_{rs} (r \neq s)$  for networks generated using SBM is 0.05, the within-community edge probabilities $ \theta_{rr}$ are given by 0.6, 0.55, 0.5, 0.45, 0.4, 0.35, 0.3, 0.25, 0.2, 0.15, respectively. The sizes of two communities are unequal, such as 400 $\&$ 600 (i.e., $n=1000, \pi_1=0.4, \pi_2=0.6$) in Fig.1(a), 200 $\&$ 300 in Fig.1(b) and 30 $\&$ 50 in Fig.1(c).

GN network is a special case of SBM network, where there are 128 nodes with 4 communities, and $\pi_r=0.25 \; (r=1, 2, 3, 4)$. On average, there are 16 neighbors for each node with $Z_{in}$ ones in its own community and $Z_{out}$ ones in the rest, i.e., $Z_{in}$ + $Z_{out}$=16, and $ \theta_{rr}=\dfrac{Z_{in}}{32}, \theta_{rs} =\dfrac{Z_{out}}{96} \; (r \neq s, r, s=1,2, \cdots, c)$.

2. LFR network

LFR network can uncover the characteristics of real networks that the  distributions of degrees and community sizes are power laws with exponents $\gamma$ and $\beta$, respectively. LFR network is generated using parameters $n$ (the number of nodes), $\mu $ (the mixing parameter), $k$ (the average degree of nodes), $maxk$ (the maximum degree of nodes), $minc$ (the minimum for the community sizes) , $maxc$ (the maximum for the community sizes). The strength of network structure is controlled by mixing parameter $\mu$ ($\mu \in [0,1]$), which is the fraction that a node connects to the ones in other communities.

In this paper,  $\gamma$ and $\beta$ are set to be 2 and 1, respectively, and the other parameters are given in the following experiments.

\subsection{Equivalence test on synthetic networks}
In this section, we validate the equivalence relations between $Q$ and NMF, and $D$ and NMF.

Firstly, we test the approximate equivalence relation  between $Q$ and NMF introduced in Setc.II on SBM networks and LFR networks. The results are illustrated in Fig.1 (on SBM networks) and Fig.2 (on LFR networks), from which one can see that, when $n$ is large, all of the points are on a straight line. However, with the decrease of $n$, gradually these points are not on a straight line. These phenomena are consistent with our conclusions in Sect.II: maximizing $Q$ and minimizing the object function of NMF with Frobenius norm are equivalent when $n$ is moderately large.

\begin{figure}[htbp]
	\centering
	\includegraphics[width=16.5cm]{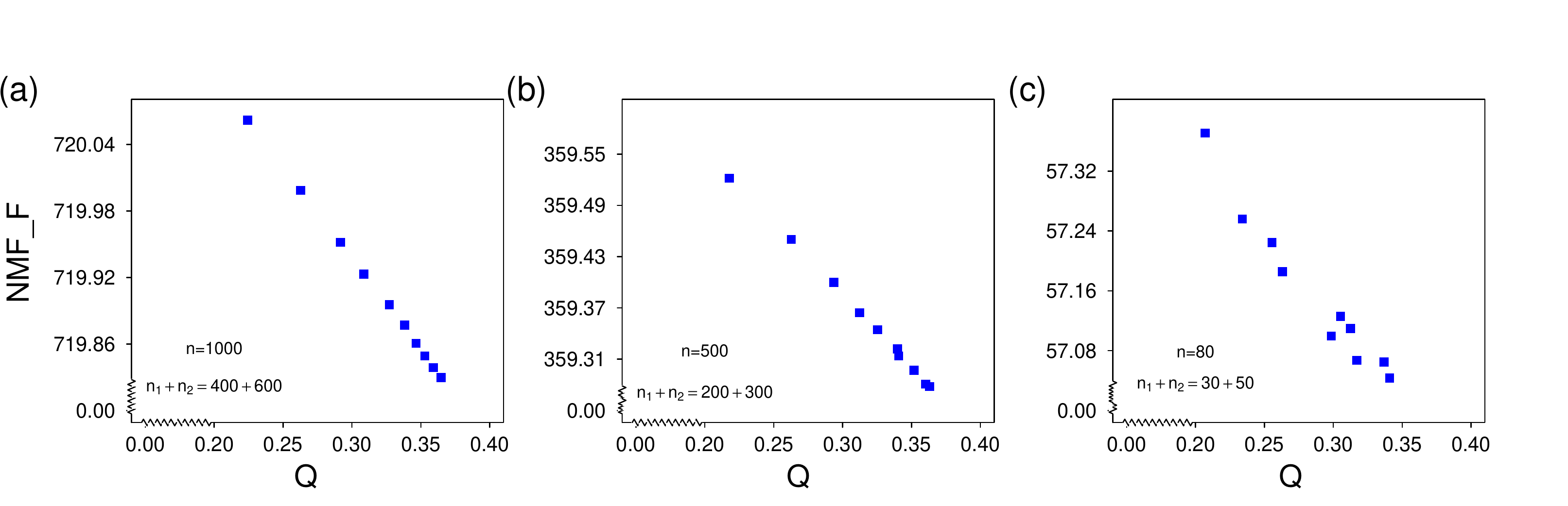}
	\caption{Approximate equivalence between maximizing $Q$ and NMF with Frobenius norm (NMF\_F) on SBM networks. Each point corresponds to a SBM network. All networks were generated by the stochastic block model with two communities. $n$ is the number of nodes. $n_1$ and $n_2$ are the sizes of two communities, respectively. The between-community edge probability for networks is 0.05, the within-community edge probabilities (corresponding to the points from left to right) are given as 0.6, 0.55, 0.5, 0.45, 0.4, 0.35, 0.3, 0.25, 0.2, 0.15, respectively. (a) The networks contain 1000 nodes. The sizes of two communities are 400 and 600, respectively. (b) The networks contain 500 nodes. The sizes of two communities are 200 and 300, respectively. (c) The networks contain 80 nodes. The sizes of two communities are 30 and 50, respectively.}\label{fig:K1}
\end{figure}

\begin{figure}[htbp]
	\centering
	\includegraphics[width=16.5cm]{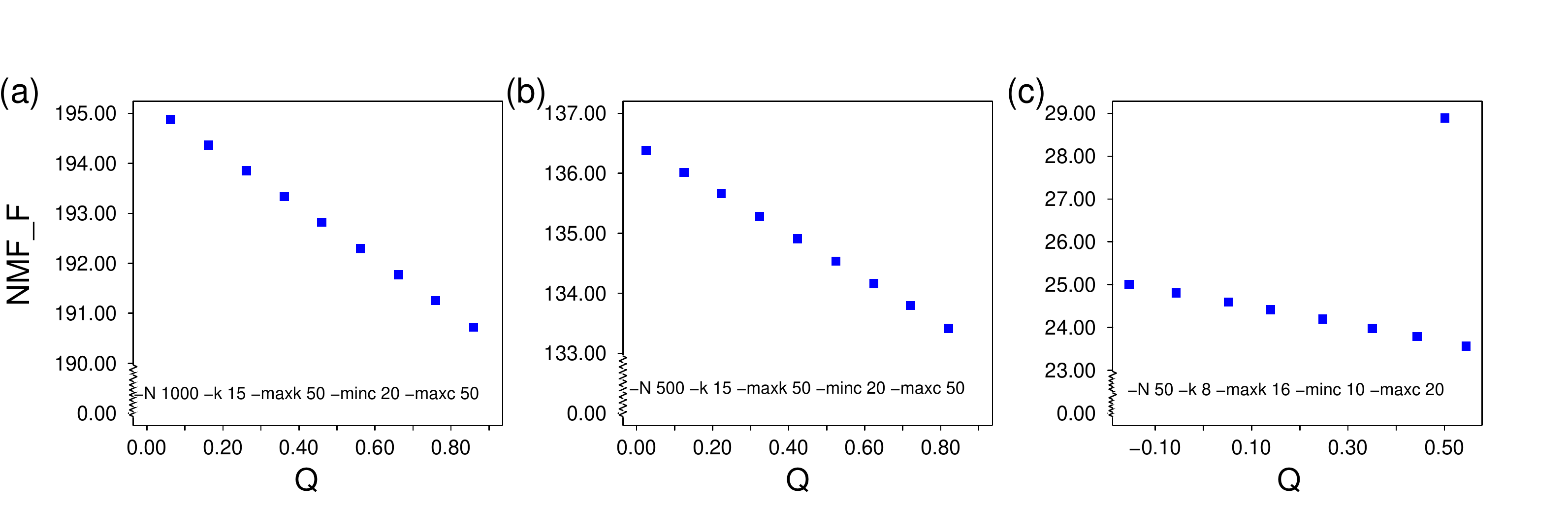}
	\caption{Approximate equivalence between maximizing $Q$ and NMF with Frobenius norm (NMF\_F) on LFR networks. Each point corresponds to a LFR network, and some of the parameters generated networks are shown in the illustration. The nine points (from left to right) correspond to mixing parameters  $ \mu=0.1, 0.2, \cdots, 0.9$, respectively.}\label{fig:K2}
\end{figure}

Secondly, we test the equivalence relation between $D$ and NMF in Sect.III on SBM networks and LFR networks. The results are illustrated in Fig.3 (on SBM networks) and Fig.4 (on LFR networks),
from which one can observe that, all of the points are on a straight line, and different values of $\sigma$ in Eq.(14) do not affect the results.

\begin{figure}[htbp]
	\centering
	\includegraphics[width=14.5cm]{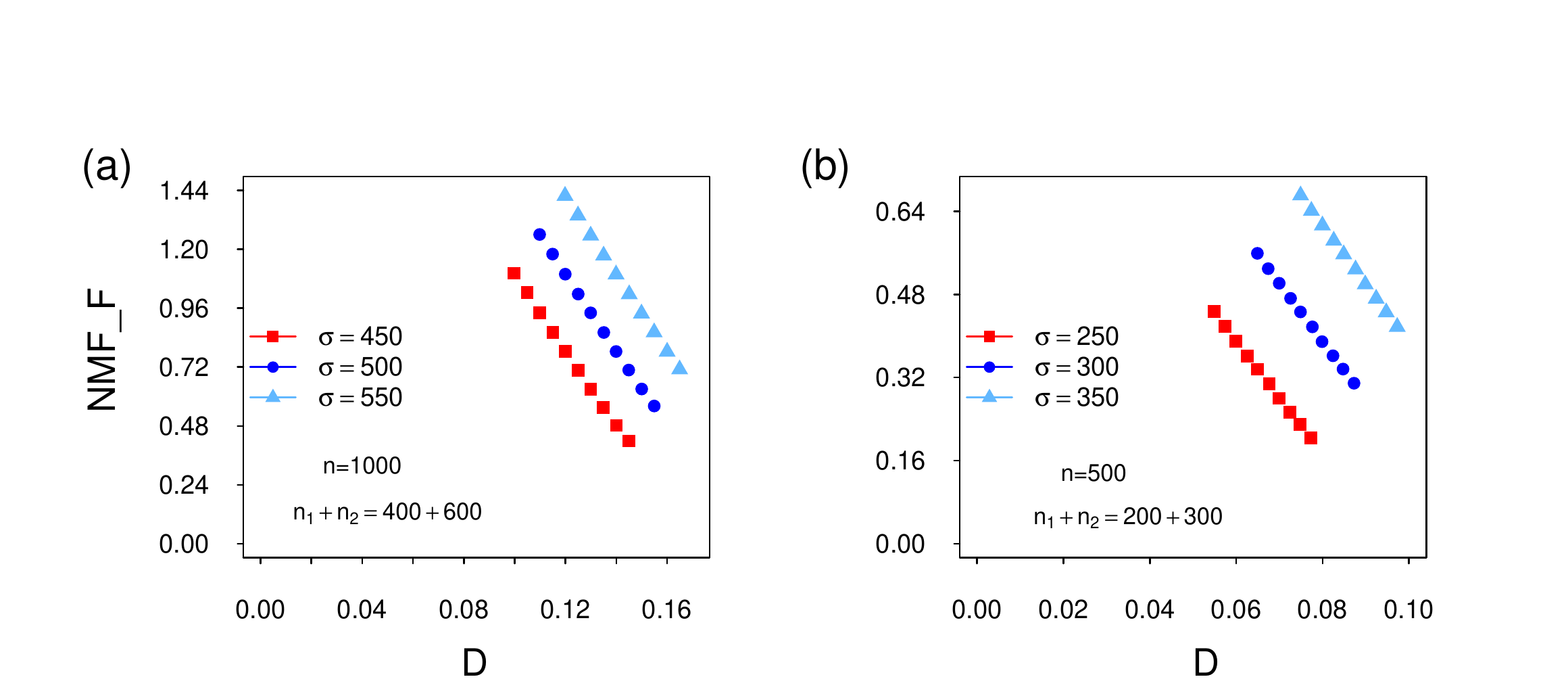}
	\caption{Equivalence relation between maximizing $D$ and NMF with Frobenius norm (NMF\_F) on SBM networks. Each point corresponds to a SBM network. Parameter settings are identical with that in Fig. \ref{fig:K1}.}\label{fig:K3}
\end{figure}

\begin{figure}[htbp]
	\centering
	\includegraphics[width=16.5cm]{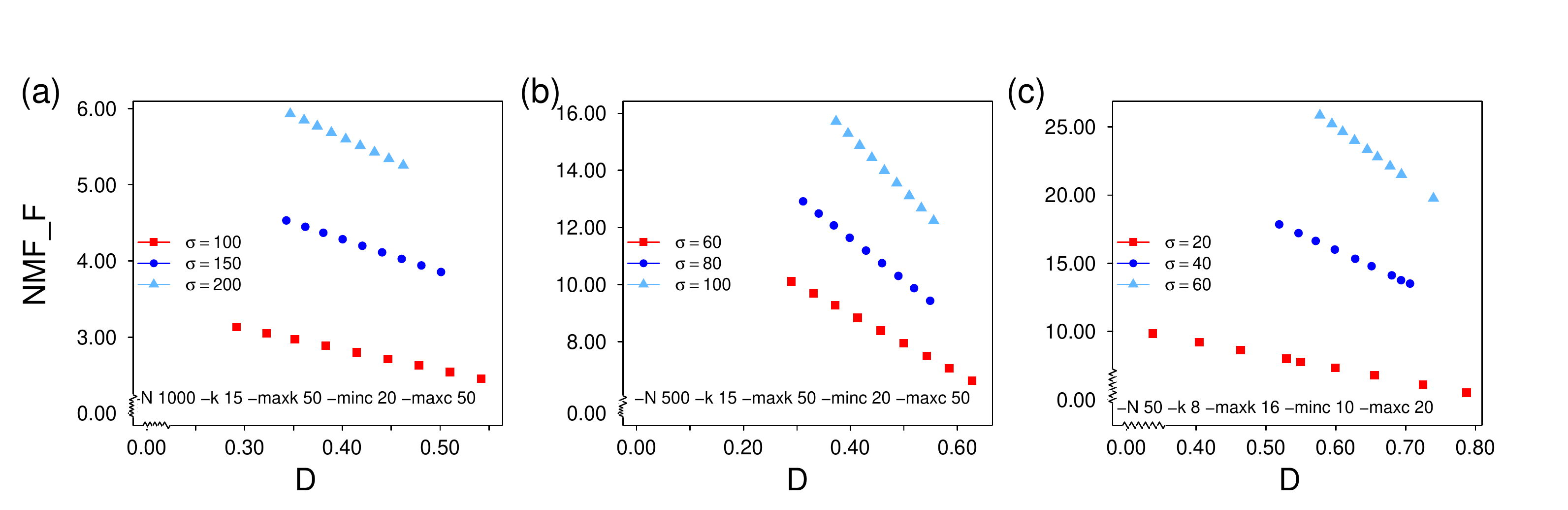}
	\caption{Equivalence relation between maximizing $D$ and NMF with Frobenius norm (NMF\_F) on LFR networks. Parameter settings are identical with that in Fig. \ref{fig:K2}.}\label{fig:K4}
\end{figure}

Finally, we test the equivalence relation between $Q$ and NMF in Sect.IV.  Since this equivalence has the constraint on node degrees, i.e., all nodes have the same degree, we only test the equivalence on GN networks. The results are illustrated in Fig.5, from which one can see that: (1) for (a), all pairs of points are in a straight line, which illustrates the equivalence between $Q$ \& NMF in Sect.IV is reasonable; (2) for (b) and (c), all pairs of points are on a straight line, and different values of parameter $\gamma$ and $r$ cannot affect the equivalence, which means that the equivalence of $Q^{RB} $ \& NMF and $Q^{AFG} $ \& NMF is reasonable, respectively. In brief, these illustrations are consistent with our conclusions in Sect.IV.

\begin{figure}[htbp]
	\centering
	\includegraphics[width=16.5cm]{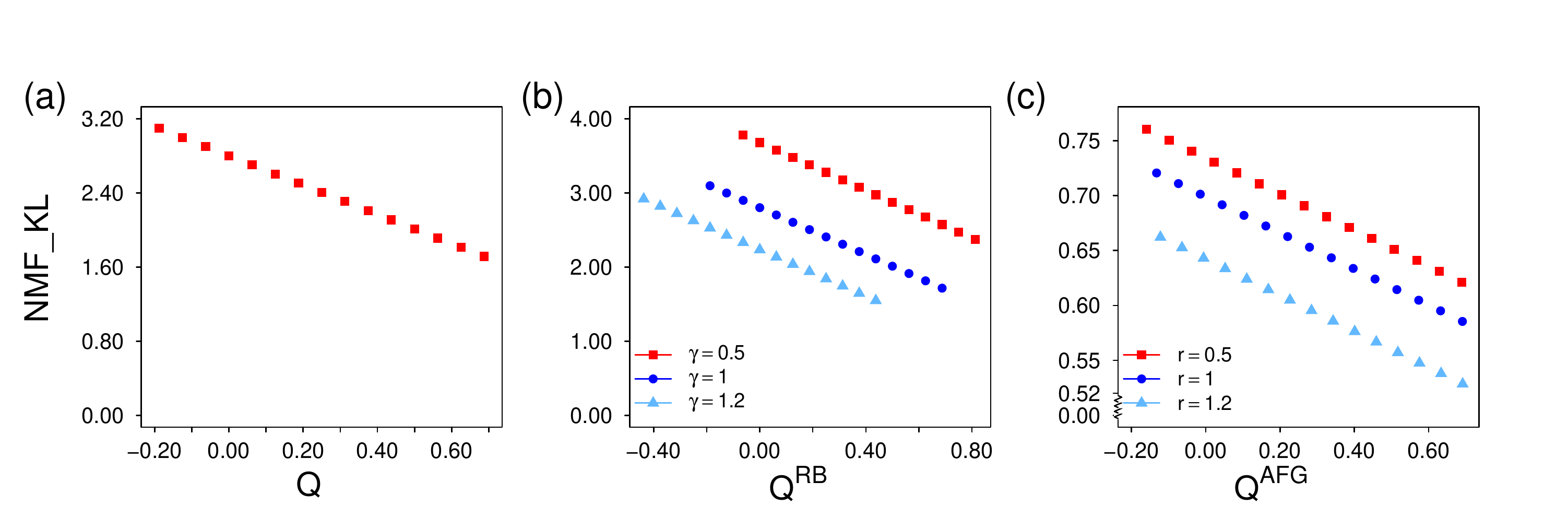}
	\caption{Equivalence relation between minimizing Q and NMF with KL divergence (NMF\_KL) on GN networks. The 15 points (from left to right) corresponds to $Z_{out}=1, 2, \cdots, 15$, respectively.}\label{fig:K5}
\end{figure}

\subsection{Comparison of algorithm effectiveness on  GN networks and LFR networks}
In this section, we design the algorithms for model (7) and (15), denoted by Q\_NMF and D\_NMF, respectively, which can be summarized in Algorithm 1. The only difference is that they use different objective functions. In addition,  $Q$ and $D$ can also be reformulated as the traces of matrices called modularity Laplacian with nonnegative relaxation for community detection \cite{jiang2012modularity}, denoted by Q\_NR and D\_NR, respectively, and $Q$ function itself is often used as objective function of optimization for community detection. We compare the efficiencies of these five algorithms on GN networks and LFR networks, and use normalized mutual information (NMI) \cite{37Strehl2003Cluster} to evaluate the quality of the results, i.e.,
\[
NMI(V^{\left( T \right)} ,V^{\left( I \right)} ) = \frac{{\sum\limits_{i = 1}^c {\sum\limits_{j = 1}^c {n_{ij} \log \frac{{nn_{ij} }}{{n_i^{V^{\left( T \right)} } n_j^{V^{\left( I \right)} } }}} } }}{{\sqrt {\left( {\sum\limits_{i = 1}^c {n_i^{V^{\left( T \right)} } \log \frac{{n_i^{V^{\left( T \right)} } }}{n}} } \right)\left( {\sum\limits_{j = 1}^c {n_j^{V^{\left( I \right)} } \log \frac{{n_j^{V^{\left( I \right)} } }}{n}} } \right)} }}
\]
where $V^{(T)} = \left( {V_{1}^{(T)} ,V_{2}^{(T)} , \cdots ,V_{c}^{(T)} } \right)$ are true communities in a network, $V^{(I)} = \left( {V_{1}^{(I)} ,V_{2}^{(I)} , \cdots ,V_{c}^{(I)} } \right)$ are infered communities; $c$ is the number of  communities contained in a network; $n$ is the number of nodes; $n_{ij}$ is the number of nodes in the ground truth community $V_i^{(T)}$ that are assigned to the computed community $V_{j}^{(I)}$;  ${n_i^{V^{\left( T \right)} } }$ is the number of nodes in the ground truth community $V_i^{(T)}$; ${n_j^{V^{\left( I \right)} } }$ is the number of nodes in the computed community $V_{j}^{(I)}$. A larger $NMI$ means a better partition.

\begin{algorithm}[H]
	\caption{Minimizing $ \left\| {W - SS^T } \right\|_F^2$ with respect to $S$}
	\label{Al:01}
	\begin{algorithmic}[1]
		\REQUIRE $A, c$, iter
		\ENSURE $S$
		\STATE Calculating $W$ by $A$.
		\STATE Initializing $S $.
		\FOR{$t=1:\mbox{iter}$}\vspace{2mm}
		\STATE
		$S : = S\odot \frac{{\left( {WS} \right) }}{{\left( {SS^T S} \right) }}$
		\vspace{2mm}
		\ENDFOR
	\end{algorithmic}
\end{algorithm}

\begin{figure}[htbp]
	\centering
	\includegraphics[width=14.5cm]{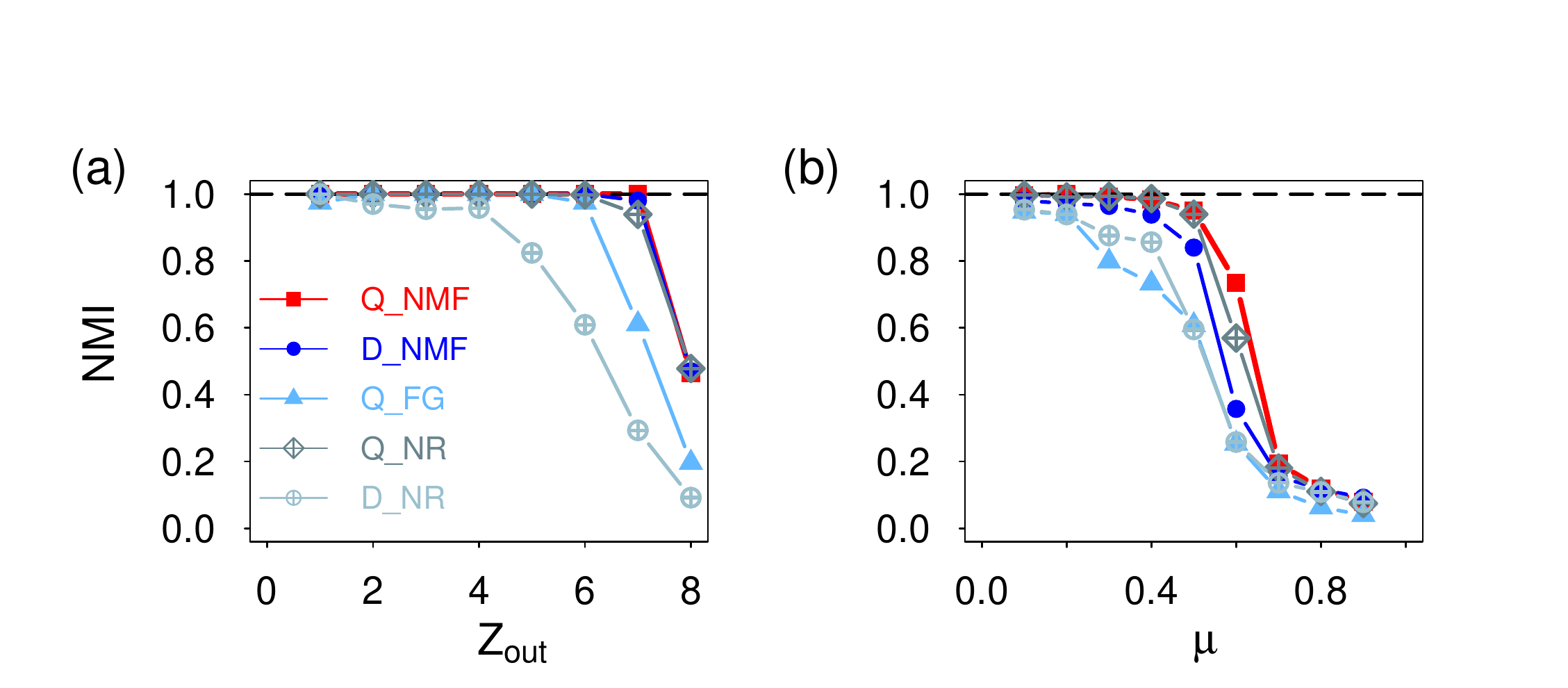}
	\caption{Comparison of the effectiveness of the five methods on GN (a) and LFR networks (b).
Q\_NMF and D\_NMF stand for multiplicative update rules for $Q$ and $D$, respectively (i.e., Eq.(7) and Eq.(15)). Q\_FG stands for fast greedy algorithm for Q. Q\_NR and D\_NR stand for nonnegative relaxation method for $Q$ and $D$, respectively. Each point is calculated from the average of 10 runs. For each run, the iterations of Algorithm 1 are 500.}\label{fig:K6}
\end{figure}

Fig. \ref{fig:K6} shows the averaged NMI calculated by the five algorithms on (a) GN networks and (b) LFR networks, from which one can see that the winner of these five algorithms is Q\_NMF, which is proposed in this paper.

\section{Conclusions}
In this paper, we establish the (approximate) relations between $Q$ optimization and NMF, and $D$ optimization and NMF. The effectiveness of the algorithms employed by $Q$, $D$ and NMF are compared demonstrating that the multiplicative update rules proposed in this paper are more effective. There are several interesting problems for future work including designing more effective algorithms for community detection, developing more reasonable criteria for evaluation of community structures in networks, and introducing the ideas of modularity $Q$ into general clustering analysis.


\bibliography{bipartite}
\end{document}